\documentclass[showpacs,prl,twocolumn]{revtex4}
\usepackage{graphicx}
\usepackage{epsfig}
\usepackage{dcolumn}
\usepackage{amsfonts}
\usepackage{amsmath}
\usepackage{amssymb}
\usepackage{bm}
\begin{document}
\title{Scaling behavior of linear polymers in disordered media}
\author{Hans-Karl Janssen}
\affiliation{Institut f\"{u}r Theoretische Physik III, Heinrich-Heine-Universit\"{a}t,
40225 D\"{u}sseldorf, Germany }
\author{Olaf Stenull}
\affiliation{Fachbereich Physik, Universit\"{a}t Duisburg-Essen, Campus Duisburg, 47048 Duisburg,
Germany }
\date{\today}

\begin{abstract}
Folklore has, that the universal scaling properties of linear polymers in disordered
media are well described by the statistics of self-avoiding walks (SAWs) on
percolation clusters and their critical exponent $\nu_{\text{SAW}}$, with SAW implicitly referring to \emph{average} SAW. Hitherto, static averaging has been commonly used, e.g.\ in numerical simulations, to determine what the \emph{average} SAW is. We assert that only kinetic, rather
than static, averaging can lead to asymptotic scaling behavior and corroborate
our assertion by heuristic arguments and a renormalizable field theory. Moreover,
we calculate to two-loop order $\nu_{\text{SAW}}$, the exponent $\nu
_{\text{max}}$ for the longest SAW, and a new family of multifractal exponents
$\nu^{(\alpha)}$.

\end{abstract}
\maketitle

%\keywords{Percolation, self-avoiding walks, linear polymers, field theory, renormalization group, critical exponents}

In the past twenty years, the critical behavior of polymers in disordered media
has generated a great deal of interest (for a recent review see~\cite{Chak05}). The problem is relevant in a vast range of different fields. To name a
prominent example, the transport properties of polymeric chains in porous
media might be exploitable commercially to enhance oil recovery. It has long been known, that
polymers in disordered media are well modelled by self-avoiding walks (SAWs)
on percolation clusters. The term SAW usually refers implicitly to {\em average} SAW.
Despite of many ideas put forward and extensive numerical
efforts, the critical behavior of polymers in disordered media is still far
from being completely understood. The most unsettling problems are, perhaps,
on the analytical side that the only existing field theoretic model for
studying {\em average} SAWs, the Meir-Harris (MH) model \cite{MeHa89}, has trouble with renormalizability \cite{DoMa91,FeBlFoHo04} and that, on the numerical side, simulations lead to widespread results for the scaling exponent $\nu_{\text{SAW}}$ describing the mean length of {\em average} SAWs, see~\cite{Chak05}.

A conceptual subtlety, that apparently has not been appreciated much hitherto,
is the precise meaning of \emph{average} SAW. Viz.\ there are, essentially,
two qualitatively different ways of averaging over all SAWs between two
connected sites for a given random configuration of a diluted lattice, one
being static and the other being kinetic. In this letter, we conjecture that
the statistics of linear polymers in disordered media has no asymptotic
scaling limit, when static averaging is used. Since it is static averaging
that has been commonly employed in numerical work, many simulations might have
suffered from this non-scaling behavior, which could explain the discrepancies between the
numerical results for $\nu_{\text{SAW}}$. In the following, we first
corroborate our conjecture by heuristic arguments. Then, we resort to
renormalized field theory. It turns out that, for achieving renormalizability, one has
to use kinetic averaging; static averaging leads to non-renormalizability. We then employ our field theory to calculate $\nu_{\text{SAW}}$, the corresponding scaling exponents for the
shortest and longest SAWs and an entire family of multifractal exponents to
two-loop order. Finally, we discuss the connection of kinetic averaging and
the MH model.

First, let us define what we mean by static and kinetic averaging. In the framework of field theory, it is most convenient to model linear polymers in disordered media as SAWs between two connected
sites $x$ and $y$ of a diluted lattice, were bonds are occupied with a
probability $p$, and to focus on the the length $L(x,y)$ of a SAW (a random number proportional to the number of monomers of the corresponding polymer) rather than the Euklidian distance $\left\vert \mathbf{x}-\mathbf{y}\right\vert$ of its endpoints~\cite{MeHa89}. First, let us consider one given random configuration $\mathfrak{C}$ of the diluted lattice. Averaging over all of SAWs belonging to the bundle $\mathcal{B}(x,y;\mathfrak{C})$ of SAWs directed from $y$ to $x$ yields the mean length
\begin{equation}
\langle L(x,y)\rangle_{\mathfrak{C}}=K\frac{\partial}{\partial K}\ln\Big(\sum_{\gamma\in\mathcal{B}
(x,y;\mathfrak{C})}p(\gamma)K^{L(\gamma)}\Big)\,,
\end{equation}
where $L(\gamma)$ is the length of $\gamma$, $p(\gamma)$, with
$\sum_{\gamma}p(\gamma)=1$, is a weight factor that depends on the averaging
procedure, and  $K$ is the fugacity. Static averaging means that one simply uses $p(\gamma)\propto 1$. Kinetic averaging, on the other
hand, means that a SAW $\gamma$ earns a factor $1/z$ contributing to
$p(\gamma)$ at each ramification where $z-1$ other SAWs from the bundle
$\mathcal{B}(x,y;\mathfrak{C})$ split off. Experimentally relevant, however,
is not $\langle L(x,y)\rangle_{\mathfrak{C}}$ but rather its average $\left[\cdots\right]_{p}$ over all
configurations $\mathfrak{C}$ at fixed $p$ subject to the constraint, that $x$
and $y$ are connected. This average is expected to exhibit scaling behavior,
\begin{equation}
M(x,y)=\left[\langle L(x,y)\rangle_{\mathfrak{C}}\right]_{p}
\sim\left\vert \mathbf{x}-\mathbf{y}\right\vert ^{1/\nu_{\text{SAW}}},
\label{length-scal}
\end{equation}
at a critical value $K_{c}$ of the fugacity.

As we will demonstrate, SAWs on a percolation cluster are not merely
standard fractals. Rather, they are multifractals. In order to capture this
multifractality, we define the bond-weights $m_{b}=\sum_{\gamma\in
\mathcal{B}(x,y;\mathcal{C})}\chi_{b}(\gamma)p(\gamma)\leq1 $, where $\chi
_{b}(\gamma)$ is one if the bond $b$ belongs to the SAW $\gamma$ and zero
otherwise, and we introduce the multifractal moments
\begin{equation}
L^{(\alpha)}(x,y)=\sum_{b}s_{b}m_{b}^{\alpha} \label{MultiMom}
\end{equation}
with $s_{b}$ being the length of bond $b$. We will show that the scaling
behavior of their quenched averages,
\begin{equation}
M^{(\alpha)}(x,y)=\big[L^{(\alpha)}(x,y) \big]_{p}
\sim\left\vert \mathbf{x}-\mathbf{y}\right\vert ^{1/\nu^{(\alpha)}} ,
\label{MultiFrak}%
\end{equation}
is characterized by multifractal exponents $\nu^{(\alpha)}$ satisfying
$\nu^{(0)}=1/D_{bb}$, $\nu^{(1)}=\nu_{SAW}$, and $\nu^{(\infty)}=\nu$, where
$D_{bb}$ is the fractal dimension of the backbone and $\nu$ is the percolation
correlation length exponent.

\begin{figure}[ptb]
%\begin{center}
\includegraphics[width=3.0cm]{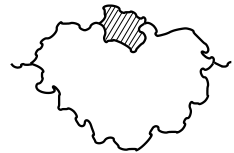}\caption{Percolation cluster in
the node-link-blob picture.}
\label{nodeLinkBlobCluster}
\end{figure}

It is well known \cite{Pe84_KrLy85_Pi85} that in a non-random medium ($p=1$)
the exponent $\nu_{SAW}$ is the same for static and kinetic averaging. This may be
not the case in a random medium, at least at the percolation point, and static
averaging does not lead to a scaling law like Eq.~(\ref{length-scal}).
Heuristically, this can be understood by employing the node-link-blob picture
of percolation clusters in which a percolation cluster connecting two terminal
points, which is generically very inhomogeneous and asymmetric, can be envisaged as two nodes linked by tortuous ribbons that may contain blobs consisting of many short links in their interior.
We will now use this picture to demonstrate that
static averaging is unstable against coarse graining and that it therefore can
not be expected to produce the correct asymptotic scaling behavior. 
Let us for simplicity consider the cluster sketched in Fig.~\ref{nodeLinkBlobCluster} that features two links, one with and the other without a blob. With static averaging the (upper) link
with the blob acquires a much larger weight then the other (lower) one even if
it is much shorter than the link with the blob. Then, the statistics of the
mean length is dominated by the short upper link with its many different SAWs
induced by the blob. However, the weights change drastically upon coarse
graining. Suppose we have some coarse graining procedure that culminates in
condensing the blob into a single bond. After that, both links have the same
weight. However, the lower one, since it is longer, now dominates the
statistics. This demonstrates the instability of the weights of static
averaging under real space renormalization as the group generated by repeated
coarse graining. In contrast, kinetic averaging does assign the same weight to
both links independent of the blob. Thus, kinetic averaging is stable under
coarse graining even in a strongly inhomogeneous disordered medium.
%%%%%%%%%%%%%%%%%%%%%%%%%%%
%%%%%%%%%%%%%%%%%%%%%%%%%%%

To fortify our arguments, we now turn to renormalized field theory. We will
propose a theory for calculating $\nu_{SAW}$, as well as the entire family
$\nu^{(\alpha)}$, that is renormalizable, provided that kinetic averaging is
used. This theory is based on the nonlinear random resistor network (nRRN),
where any bond on a $d$ dimensional lattice is occupied with a resistor with
probability $p$ or respectively empty with probability $1-p$. Our theory, is
motivated by the well know fact, that the shortest and the longest SAW (the
former is also known as the chemical path) can be extracted from the
nRRN~\cite{KeSt82/84} and its field theoretic formulation, the Harris
model~\cite{Ha87}, by considering specific limits of the nonlinearity $r$ of
the generalized Ohm's law governing the bond resistors,
\begin{equation}
V_{j}-V_{i}=\rho_{(ij)}\left\vert I_{i,j}\right\vert ^{r}\,\operatorname{sign}{I_{i,j}} \, ,
\label{Ohm-V}%
\end{equation}
where $V_{i}$ is the voltage at lattice site $i$, $\rho_{(ij)}$ is the
resistance of bond $(ij)$ and $I_{i,j}$ is the current flowing through that
bond. As shown rigorously by Blumenfeld \emph{et al}.~\cite{BMHA85/86}, the
shortest and the longest SAW correspond to $r \to+0$ and $r \to-0$
respectively. Evidently, $M(x,y)$ must lie in between the average length of
the shortest and the longest SAW, which are, of course, very different. Since
$M(x,y)$ sits somewhere in this discontinuity at $r=0$, it is not known how to
extract it from the nRRN by a limit taking. Therefore, we propose
here to study the \emph{average} SAW by using our real-world
interpretation~\cite{JaStOe99,JaSt00,St00,StJa00/01}, in which the
Feynman diagrams for the nRRN are viewed as being resistor networks themselves.
The idea is to put SAWs  on these diagrams. That this idea is fruitful can be
checked explicitly at the instance of the chemical path. Our approach
reproduces to two-loop order the corresponding exponent $\nu_{\text{min}}$
well establish from dynamical percolation theory~\cite{Ja85}. 

Our field theory is based on the Harris model as described by the Hamiltonian
\begin{equation}
\mathcal{H}=\int d^{d}x\sum_{\vec{\theta}}\Big\{ \frac{\tau}{2}\varphi^{2}
+\frac{1}{2}(\nabla\varphi)^{2}+\frac{w}{2}\varphi\,\bigl(-\vec{\partial
}_{\theta}\bigr)^{r+1}\varphi+\frac{g}{6}\varphi^{3}\Big\} ,
\label{Hamiltonian}
\end{equation}
where $\vec{\theta}$ is a replicated discretized voltage taking on $(2N)^{D}$
values on a $D$-dimensional torus: $\vec{\theta} = \frac{\pi}{N} (n_1, \cdots, n_D)$ with $n_i = -N +1, -N+2, \cdots, N - 1, N$. $\varphi= \varphi(\mathbf{x},\vec{\theta})$ is the order parameter
field, a continuum analog of a Potts spin. It transforms according to the {\em one} 
irreducible representation of the symmetric (permutation) group $S_{(2N)^{D}}$ and thus, the model features only a single coupling constant, $g$. $\tau$ and
$w$ are strongly relevant critical control parameters. The scaling behavior of
SAWs is associated with the renormalization of $w$ in the replica limit
$D\to0$. For details on the Harris model, we refer
to~\cite{MeHa89,JaStOe99,JaSt00}. The diagrammatic perturbation theory of the
Harris model can be formulated in such a way that the Feynman diagrams
resemble real RRNs. In this approach, which we refer to as real
world interpretation, the diagrams feature conducting propagators
corresponding to occupied, conducting bonds and insulating propagators
corresponding to open bonds. The conducting bonds carry replica currents conjugate to the replica voltages $\vec{\theta}$. The resistance of a conducting bond is given by its Schwinger parameter~\cite{Am84_ZiJu96}. In the following, we will refer only to these very basic aspects of the real world interpretation which will be sufficient to follow the main line of argument. For further details on the real world interpretation, see Refs.~\cite{JaStOe99,JaSt00,St00,StJa00/01}.

Now we employ the real world interpretation to study the \emph{average} SAW.
As an example, let us consider the two-loop diagram that resembles the
node-link-blob cluster in Fig.~\ref{nodeLinkBlobCluster} and where all
internal propagators are conducting. For determining the contribution of this
diagram to $M(x,y)$, the essential step in our approach is to find out the
length $L$ of an average SAW on that diagram. We can either apply the static
or the kinetic rule to calculate $L$ according to Eq.~(\ref{MultiMom}) with
$\alpha=1$, where now bond $b$ is replaced by conducting propagator $i$ and the
Schwinger parameter~\cite{Am84_ZiJu96} $s_{i}$ is interpreted as the corresponding length.
%%%%%%%%%%%%%%%%%%%%%%%%%%%
\begin{figure}[ptb]
%\begin{center}
\includegraphics[width=2.5cm]{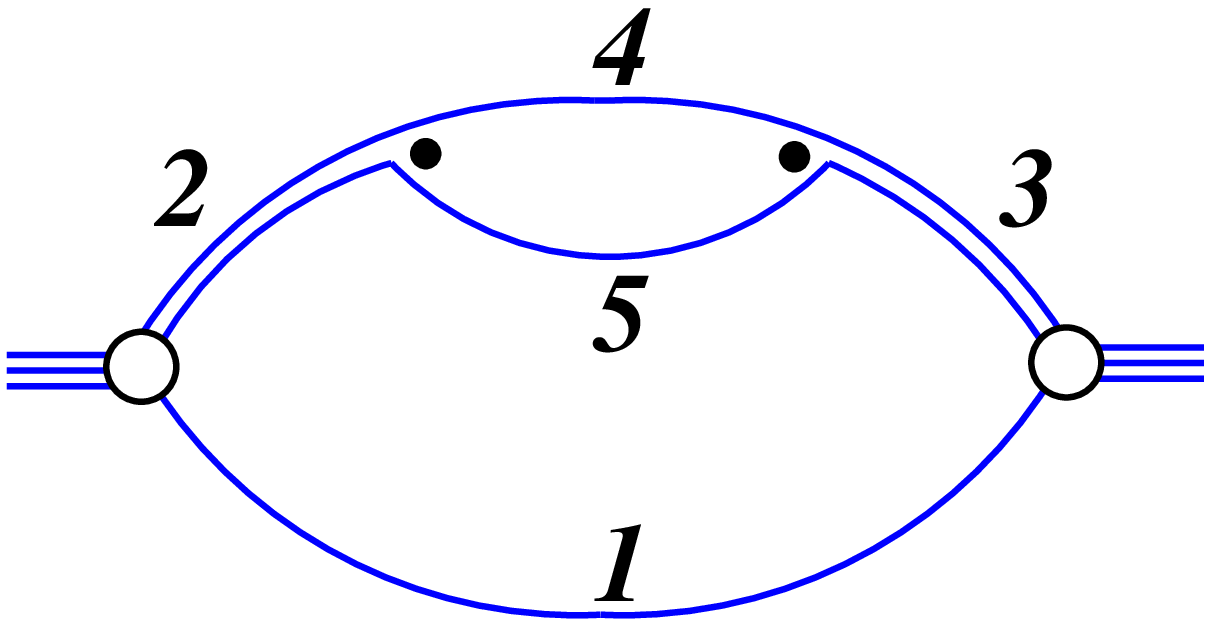} \qquad\qquad
\includegraphics[width=2.5cm]{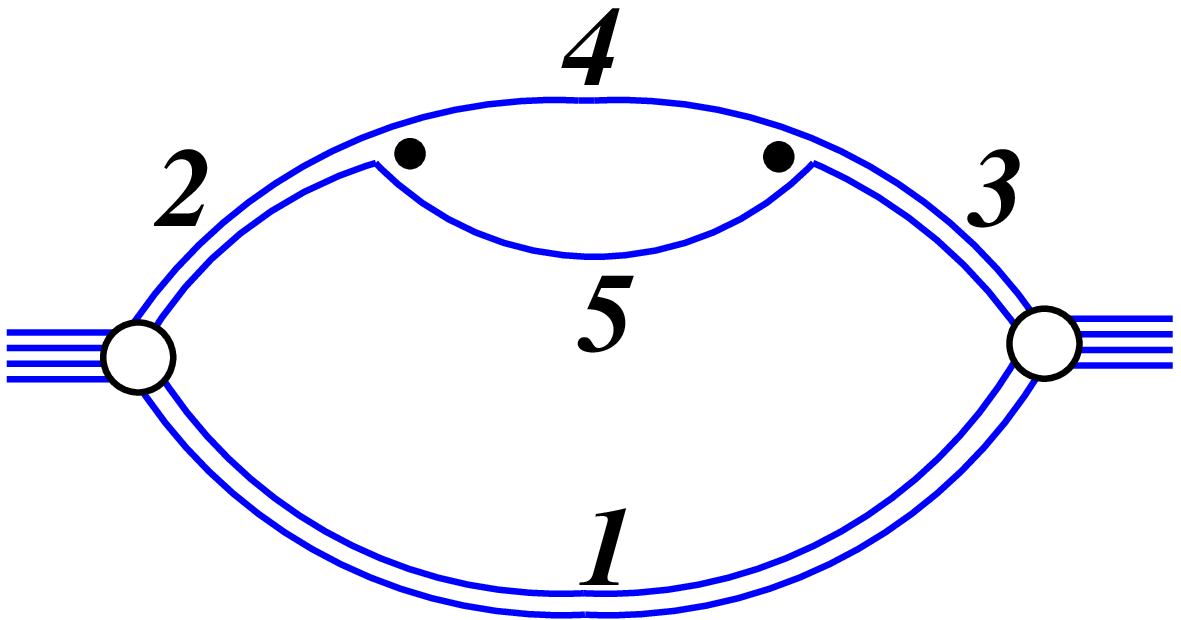}
%\label{Elemente}(hier)%
%\end{center}
\caption{(Color online) Static and kinetic rule for averaging SAWs.}
\label{Vert}
\end{figure}
%%%%%%%%%%%%%%%%%%%%%%%%%%%
As visualized in Fig.~(\ref{Vert}), static and kinetic averaging yields
different results. The static rule gives $L^{(\text{st})}=(s_{1}+s_{4}
+s_{5})/3+2(s_{2}+s_{3} )/3$, whereas the kinetic rule produces
$L^{(\text{kin})}=(s_{1}+s_{2} +s_{3})/2+(s_{4}+s_{5})/4$. The remaining steps
in calculating the diagram essentially textbook matter~\cite{Am84_ZiJu96}. It turns out, that the
static rule does not lead to a renormalizable theory. The reason is easily
shown. In the $2$-loop calculation of our example diagram, the non-primitive
divergencies arising from the sub-integrations of the $1$-loop self-energy
insertion must be cancelled through the counter-terms introduced by the
renormalization of this $1$-loop insertion. However, the weights of
$L^{(\text{st})}$ are not in conformity with the weights arising in the
corresponding $1$-loop diagram with the counterterm insertion: crunching the
insertion to a point (corresponding to $s_{4}+s_{5}\rightarrow0$) leads to
$L^{(\text{st})}=s_{1}/3+2(s_{2}+s_{3})/3$ in contrast to $L^{(\text{kin}
)}=s_{1}/2+(s_{2}+s_{3})/2$, which is equal to $L$ of the $1$-loop self-energy
diagram with a point insertion. Hence, only the weighting according to the
kinetic rule works correctly in that it leads to a cancellation of
non-primitive divergencies by one-loop counterterms. Thus, the static rule has
to be rejected on grounds of renormalizability.

\begin{figure}[ptb]
%\begin{center}
\includegraphics[width=8cm]{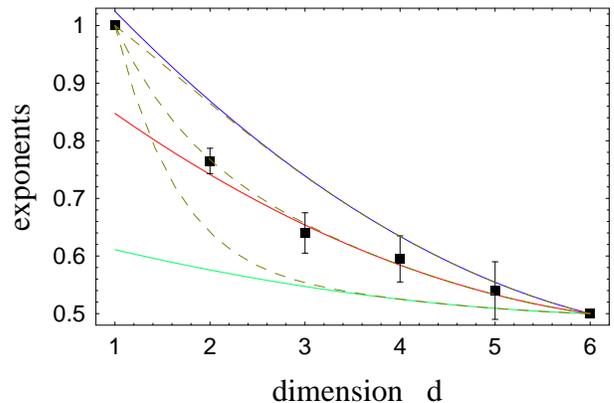}
%\label{Exponenten}(hier)
%\end{center}
\caption{(Color online) The $\varepsilon$-expansions of the exponents $\nu_{\mathrm{min}}$
(blue upper curve), $\nu_{\mathrm{SAW}}$ (red middle curve), and $\nu_{\mathrm{max}}$ (green lower curve). Possible extrapolations at low dimensions $d$ are shown by broken lines. The squares with error bars symbolize numerical results for $\nu_{\mathrm{SAW}}$ as compiled in~\cite{Chak05}. }
\label{Exp}
\end{figure}

Besides revealing the imperative of kinetic averaging, this theory yields
two-loop results for the SAW exponents $\nu_{SAW}$ and $\nu_{\text{max}}$,
which previously have been calculated (correctly) only to one-loop order
\cite{MeHa89,Ha87}, and the family $\nu^{(\alpha)}$, which is entirely new:
\begin{align}
\nu_{\mathrm{\max}}  &  =\frac{1}{2}+\frac{\varepsilon}{168}+\Big[\frac
{5365}{16464}+\frac{15}{28}\Big(\ln2-\frac{69}{70}\ln3\Big)\Big]\Big(\frac
{\varepsilon}{6}\Big)^{2}\nonumber\\
&\qquad  +O(\varepsilon^{3})\label{l-Exp-eps}
\\
\nu^{(\alpha)}  &  =\frac{1}{2}+\Big(\frac{5}{2}-\frac{3}{2^{\alpha}
}\Big)\frac{\varepsilon}{42}+\Big(\frac{589}{21}-\frac{397}{14\cdot2^{\alpha}
}+\frac{9}{4^{\alpha}}\Big)\Big(\frac{\varepsilon}{42}\Big)^{2}\nonumber\\
&  \qquad  +O(\varepsilon^{3})\,, \label{MultiFrakExp}
\end{align}
where $\varepsilon=6-d$. $\nu_{\mathrm{SAW}}$ is given by $\nu_{\mathrm{SAW}} = \nu^{(1)}$.
Our result for $\nu_{\mathrm{SAW}}$ is compared to the available numerical estimates, to our result for the longest SAW, and to the well known exponent $\nu_{\mathrm{\min}}$ for the
shortest SAW~\cite{Ja85,JaStOe99,JaSt00} in Fig.~(\ref{Exp}). The following points are worth noting: 
(i) $\nu^{(\alpha)}$ does not depend on $\alpha$ in a linear or affine fashion which implies that SAWs on percolation clusters are mulitfractal. (ii)  $\nu^{(\alpha)}$ is
in absolute agreement with the well known results for $D_{bb}$ and $\nu$ in
the cases $\alpha=0$ and $\infty$, respectively. (iii) $\nu_{\mathrm{\min}}$ and $\nu_{\mathrm{\max}}$ are not related to the family $\nu^{(\alpha)}$. (iv) the theory is renormalizable for arbitrary $\alpha$ if and only if kinetic averaging is used. 
%%%%%%%%%%%%%%%%%%%%%%%%%%%
%%%%%%%%%%%%%%%%%%%%%%%%%%%

As mentioned above, the usual framework to study \emph{average} SAWs on percolation clusters
is the MH model as described by the Hamiltonian
\begin{equation}
\mathcal{H}=\int d^{d}x\,\Big\{\sum_{k}\Psi_{k}\bigl(r_{k}-\nabla
^{2}\bigr)\Psi_{k}+\frac{g}{6}\Psi^{3}\Big\}\, . \label{MH-Hamiltonian}
\end{equation}
Here, $\Psi_{k}=\{\Psi_{k;\alpha_{1},\ldots\alpha_{k} }^{\;\;\;\;i_{1},\ldots i_{k}}(\mathbf{x})\}$, $1\leq k\leq n$ is an order parameter field conjugate to a $n$-fold replicated $m$-component Heisenberg spin with vector-indices $i_{l}$ running from $1$ to $m$ and replica indices $\alpha_{l}\in \{1,\ldots,n\}$ ordered so that $\alpha_{1}<\cdots<\alpha_{k}$. The $n$ replicas transform according to $n$ {\em different} irreducible representations of the direct product of the symmetric group $S_n$ and the orthogonal (rotation) group $SO(m)$. $r_{k}=\sum_{l}w_{l}k^{l}$, and $\Psi^{3}$ is a symbolic notation for the sum of the products of three $\Psi_{k}$ fields. Only those cubic terms
are allowed for which all pairs $(i,\alpha)$ appear exactly twice. In this
model, one can extract $\nu_{\mathrm{SAW}}$ from the renormalization of the
relevant control parameter $w_{1}$ upon taking the replica limit $n\to0$.
The MH Hamiltonian (\ref{MH-Hamiltonian}) is non-renormalizable as it stands.
One difficulty, that has been pointed out by Le~Doussal and
Machta~\cite{DoMa91} several years ago, is that the critical values
\{$r_{k}^{c}\}$ of the control parameters are different for different $k$,
i.e., the model is highly multicritical. A second problem, that to our
knowledge has not been discussed hitherto, is that the order parameter fields
$\Psi_{k}$ for different $k$ belong to different irreducible
tensor-representations of underlying symmetry group, $S_{n} \times SO(m)$, see above. Hence, strictly speaking, one needs independent coupling constants $g_{k,k^\prime,k^{\prime \prime}}$ for each product $\Psi_{k}\,\Psi_{k^\prime}\,\Psi_{k^{\prime \prime}}$ [note that this is (i) {\em not} implemented in the original MH Hamiltonian (\ref{MH-Hamiltonian}) and (ii) different in the Harris model], and the fields $\Psi_{k}$ need $k$-dependent renormalization factors~\cite{footnote}. Recently, these difficulties caused the failure of a  two-loop calculation of $\nu_{SAW}$ by von Ferber \emph{et al.} \cite{FeBlFoHo04}. 

As far as its application to the \emph{average} SAW is concerned, the
renormalizability of the MH model can be rescued by a specific interpretation
of the replica limit which has close ties to kinetic averaging. Our analysis
of the MH model (details will be given
elsewhere~\cite{JaSt06}) led to the following key findings: If the replica limit is taken after all summations
over all possible arrangements of internal replica indices of a diagram, then
the MH model reproduces static averaging and, as demonstrated above, is not
renormalizable. If, however, the replica limit is taken, in the spirit of
Ref.~\cite{Ha83b}, as early as possible, i.e.,
loop after loop, or
at least for each renormalization part,
then the MH model reproduces kinetic averaging. This is the only interpretation of replica
limit of the MH-model that leads to a renormalizable theory of SAWs in
disordered media. With this interpretation, the MH model, in particular,
produces the same result for $\nu_{\text{SAW}}$ as our
real-world interpretation and thereby provides an important consistency check
for the validity of the application of the latter to SAWs.

Closing, we would like to emphasize that our renormalization group arguments are, although certainly well founded, not rigorous in the sense of a mathematical proof since they rely on our real world interpretation of Feynman diagrams. This interpretation thrives on analogy and  there exist to date no rigorous mathematical arguments on how far its validity extends. However, given all its successes in the past, we would be surprised if it failed in describing SAWs on percolation clusters. The well known MH model, when interpreted carefully, corroborates the imperative of kinetic averaging and confirms our two-loop result for $\nu_{\text{SAW}}$.

%\section{References}


\begin{thebibliography}{99}
\bibitem{footnote}
It is not possible to construct from the $n$-fold replicated $m$-vector model a simpler model where the
order parameter fields $\Psi_{k}$ for all $k$ belong to one and the same
irreducible representation of some higher symmetry group. This is in contrast to the $n$-fold replicated $m$-state Potts model that combines to the $m^{n}$-state Potts-model, where all fields belong to
the same irreducible representation of  $S_{m^{n}}$. Therefore, in the $n$-fold replicated $m$-state Potts model, one gets away with only one coupling parameter $g$ and a unique (but $m^{n}$-dependent) renormalization factor
for all fields.
\end{thebibliography}
\end{document}